\documentclass[twocolumn,amssymb,english,superscriptaddress,letterpaper,amsmath,amsfonts,aps,pra,reprint]{revtex4-1}
\usepackage[dvipdfm]{graphicx}
\usepackage{graphicx,wrapfig}

\usepackage{natbib}
\usepackage{subfigure}
\usepackage{tabularx}
\usepackage{epsfig}
\usepackage{longtable}
\usepackage{amsfonts}
\usepackage{rotating}
\usepackage{subfigure}
\usepackage{hyperref}
\usepackage{babel}
\usepackage{currfile}
\usepackage{color}

\newcommand{\bea}{\begin{eqnarray}}
\newcommand{\eea}{\end{eqnarray}}
\newcommand{\la}{\label}
\newcommand{\be}{\begin{equation}}
\newcommand{\ee}{\end{equation}}

\def\12{\frac{1}{2}}


\usepackage{babel}

\begin{document}
\title{Effect of loss on the topological features of  dimer chain described by the extended Aubry-Andr\'e-Harper model}
\author{X. L. Zhao}
\affiliation{School of Physics and Optoelectronic Technology, Dalian
	University of Technology, Dalian 116024, China\\}
\author{Z. C. Shi}
\affiliation{Department of Physics, Fuzhou University, Fuzhou 350002, China\\}
\author{C. S. Yu}
\affiliation{School of Physics and Optoelectronic Technology, Dalian
	University of Technology, Dalian 116024, China\\}
\author{X. X. Yi}\thanks{ E-mail: yixx@nenu.edu.cn}
\affiliation{Center for Quantum Sciences and School of Physics, Northeast Normal University, Changchun 130024, China\\}
\date{\today}
\begin{abstract}
By introducing loss to one sublattice of a  dimer chain described by the extended Aubry-Andr\'e or Harper (AAH) model,
 we study the topological features including the edge states, spectrum and winding number of the chain. We find that the 
 parameter region for the system to have  real band-gap-closing is increased due to the loss, and the average displacement 
 of the single excitation can still witness  the topological features of the chain in the presence of loss. The robustness of the 
 zero energy eigenstate against four kinds of disorders is also examined. A feasible experiment setup based on coupled 
 waveguides to observe the prediction of this paper is proposed.
\end{abstract}	
\maketitle
\section{Introduction}
In 1980, Aubry and Andr\'e showed the existence of localization phase transition through a 1D tight binding quasiperiodic 
system~\cite{aips3133} described by the one dimensional (1D)   lattice model (called Aubry and Andr\'e model now). 
The model can be mapped into the 2D rectangular lattice for integer quantum Hall effect ~\cite{ppslsa68874,prb142239} 
by using Landau gauge for the magnetic field,  where the periodic character is determined by the flux quanta penetrating 
each rectangle lattice. Then the terminology `AAH model' is used widely to abbreviate  the Aubry-Andr\'e and Harper 
models.  The period for the 1D lattice is usually a trigonometrical function of length and it can be turned flexibly in principle. 
But to investigate intriguing properties, the extreme huge magnetic density is a bottleneck currently in experiments for solid 
systems~\cite{prl111185301,prl111185302,prb142239}.

With the development of topological materials, the AAH model has been explored in the view of topological 
aspects~\cite{prl108220401, prl109106402, prl109116404, pra85013638, pra90063638},  which can bridge the quantum 
Hall effect (QHE) ~\cite{prb235632,prl49405,prl713697,npb265364} and the topological insulator~\cite{rmp823045}. 
For example the 1D AAH model has the topologically protected edge states corresponding to the gapless edge states in QHE.
 Recent experiments~\cite{prl103013901,prl109106402} have realized the quasiperiodic  AAH model in optical lattices
and the signature of a localization transition~\cite{prl103013901} was observed in agreement with the theory~\cite{aips3133}.
	
The array of waveguides is a valid platform to explore topological insulators due to the developed manufacturing and designing 
technique~\cite{prl109106402,prl115040402} . However,  the loss represented by non-Hermitian terms in the system  is usually 
inevitable in practice~\cite{prl102065703,prl115040402}. It is then reasonable to consider the  influence of loss  to the 
topological properties. Considering that the AAH model may be implemented in such optical systems, we employ  in this work a 
dimer(two sites in each cell) AAH model with   loss on one of the sites in each cell,  to explore  the topological properties in terms 
of the hopping and on-site modulation phases.

Disorders exist widely in practical systems. In the original work by Aubry and Andr\'e ~\cite{aips3133}, the incommensurate
potentials mimic the disorders leading to localization transition.  Although  extensive theoretical works for the effects of disorders 
in the AAH model have been done~\cite{prl612144,prl612141,prb415544,pra80021603,prl104070601}, the study of robustness 
of topological states against these disorders is lacking. It is well known that the degenerate zero energy edge states are the edge states. 
Then we will explore  the robustness of zero energy edge states against the disorders on the chain. Four kinds of disorders are 
considered: intra-cell hopping disorders, inter-cell hopping disorders, on-site disorders and non-Hermitian disorders. We find that 
the zero energy edge states are robust against the disorders in an interval of strength except for the on-site disorders since on-site 
disorders destroy the chiral symmetry. We also find that the non-Hermitian disorders `draw' the real energy band towards zero energy.

We will explore the aforementioned  issues by considerations  that both the hopping amplitude and on-site potentials are modulated 
in the real space commensurate with the lattice in the extended AAH model~\cite{prl109116404,prb5011365}. The setups for the 
realization of this model is feasible in  coupled  waveguides with modulated lattice spacings and lattice widths, details of which are 
presented latter on.

This work is organized as follows. In section~\ref{HOP}, we introduce  a dimer chain described by the extended AAH model. In terms 
of hopping modulation phase, we study the influence of  loss on one sublattice to the topological features based on  the mean displacement 
of the single excitation on the chain. In section~\ref{ONSITE}, we examine the topological properties of the system in terms of on-site 
modulation phase.  In section~\ref{ROBUST}, we study the robustness of the zero energy edge mode against four kinds of disorders. 
In section~\ref{SETUP}, we propose  an experimental setup  to observe  the prediction  based on coupled single mode waveguides. 
Finally, we conclude  in section~\ref{SUM}.
\begin{center}
	\begin{figure}[htb!]
		\includegraphics[scale=0.39]{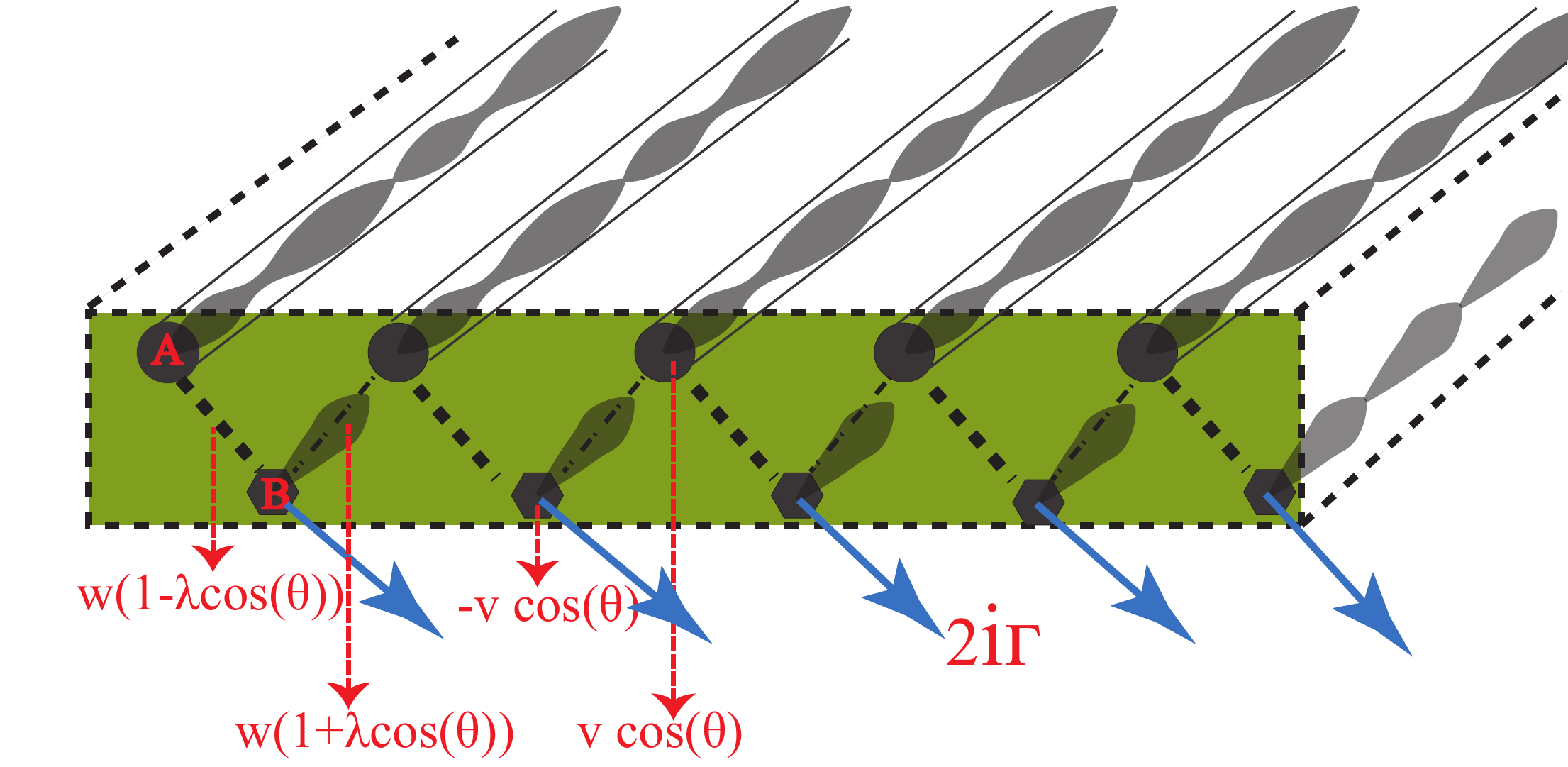}
		\vspace{0cm}
		\caption{The sketch for the extended AAH model realized in  coupled waveguides. The modulations are  functions of
			sequence of the juxtaposed waveguides. The  loss occurs  on sublattice composed of B-sites. The time evolution corresponds  
			to the propagation of light excitation along the waveguides.}
		\label{f:AAHmodelado}
	\end{figure}
\end{center}
	
\section{Topological properties in terms of hopping modulation phase}
\label{HOP}
Consider an extended 1D AAH model with modulated nearest-neighbor hopping interactions in real space
 described by the following tight-binding Hamiltonian
\bea
\label{eq:Hamiltonian}
H&=&\sum_{n=1}^{N-1} w[1+\lambda \cos(2\pi b n+\theta)] c^{\dagger}_{n+1}c_n+h.c.\nonumber\\
&&+\sum_{n=1}^{N} v \cos(2\pi b n+\theta_{v})c^{\dagger}_{n}c_n.
\eea

This 1D chain consists of $N$ sites ($n=1$, $2$, $\ldots$, $N$). Here  $c^\dagger_n$ and $c_n$ are the  polarized
fermionic creation and annihilation operators on site $n$. When we only consider a  single excitation on the chain, the operators  
$c^\dagger_n$ and $c_n$ might describe  bosons. The terms in the first line in (\ref{eq:Hamiltonian}) represent the kinetic energy 
or the nearest-neighbor hoppings. $w$ is the strength which is taken as the energy unit throughout this work and the dimensionless 
$\lambda$ indicates the modulation amplitude. The last terms describe the modulated on-site potentials where $v$ is the strength. 
$\theta$ and $\theta_v$ are the modulation phases.

When $b$ is an irrational number, the diagonal AAH model ($\lambda=0$, $v \ne 0$)  possesses a localization transition as $v$
 crosses a critical value ~\cite{aips3133,prl511198,prb408225}. In one-dimensional quasicrystals system, the topologically 
 protected boundary states  equivalent to the edge states in quantum Hall system appears when $b$ takes irrational values~\cite{prl109106402}.

We will focus on the case that $b$ is a rational number when the hopping and on-site potential modulations have the
periodicity of $1/b$ determined by the magnetic field penetrating the 2D counterpart of the 1D chain~\cite{ppslsa68874,prb142239}.
When $\lambda=0$ ($v=0$), the Hamiltonian describes the diagonal AAH model which can be derived  from  the 2D Hofstadter
 model~\cite{prl109116404,prb5011365}.

In this paper, we will  set the modulation phases of on-site potential fulfilling the condition $\theta=\theta_v+\pi$. In the experiment,
the setups can be designed  to make $\theta$  and  $\theta_v$ tunable independently such as in coupled optical waveguides systems 
since the phases is determined by modulating the spacing between the waveguides and the widths~\cite{prl109106402,prl109116404}. 
Thus it is reasonable to assume $\theta_v=\theta+\phi$ where $\phi$ is independent of $\theta$. We will focus on the topological 
properties  of this model in terms of the two modulation phases $\theta$ and $\phi$ in the following.

To simplify the problem,  we consider  $b=1/2$ case when the odd and even sites feel different commensurate hopping and on-site potentials.
Then we denote the annihilation operators for the odd sites (A-sites) as $\hat{a}$ and  $\hat{b}$ for even sites (B-sites) and treat an odd-even
 combination of the sites on the original chain as one cell, namely, the chain of dimer. In order to investigate the influence
 of loss to the topological properties of this system, we introduce non-Hermitian terms to B-sites with strength $i 2 \Gamma$, see 
 Fig.\ref{f:AAHmodelado}. The Hamiltonian reads,

\bea
H&=&H_{odd}+H_{even},\nonumber\\
H_{odd}&=&\sum_{n:odd} [1-\lambda \cos(\theta)] b^{\dagger}_{n}a_n\nonumber\\
&&+h.c-v \cos(\theta_{v}) a^{\dagger}_na_n, \nonumber\\
H_{even}&=&\sum_{n:even} [1+\lambda \cos(\theta)] a_{n+1}^{\dagger}b_n\nonumber\\
&&+h.c+(v \cos(\theta_{v})-2 i \Gamma)b^{\dagger}_nb_n.
\la{eq:Hamiltonian2}
\eea
\begin{center}
\begin{figure*}
\centering
{\includegraphics[scale=0.49]{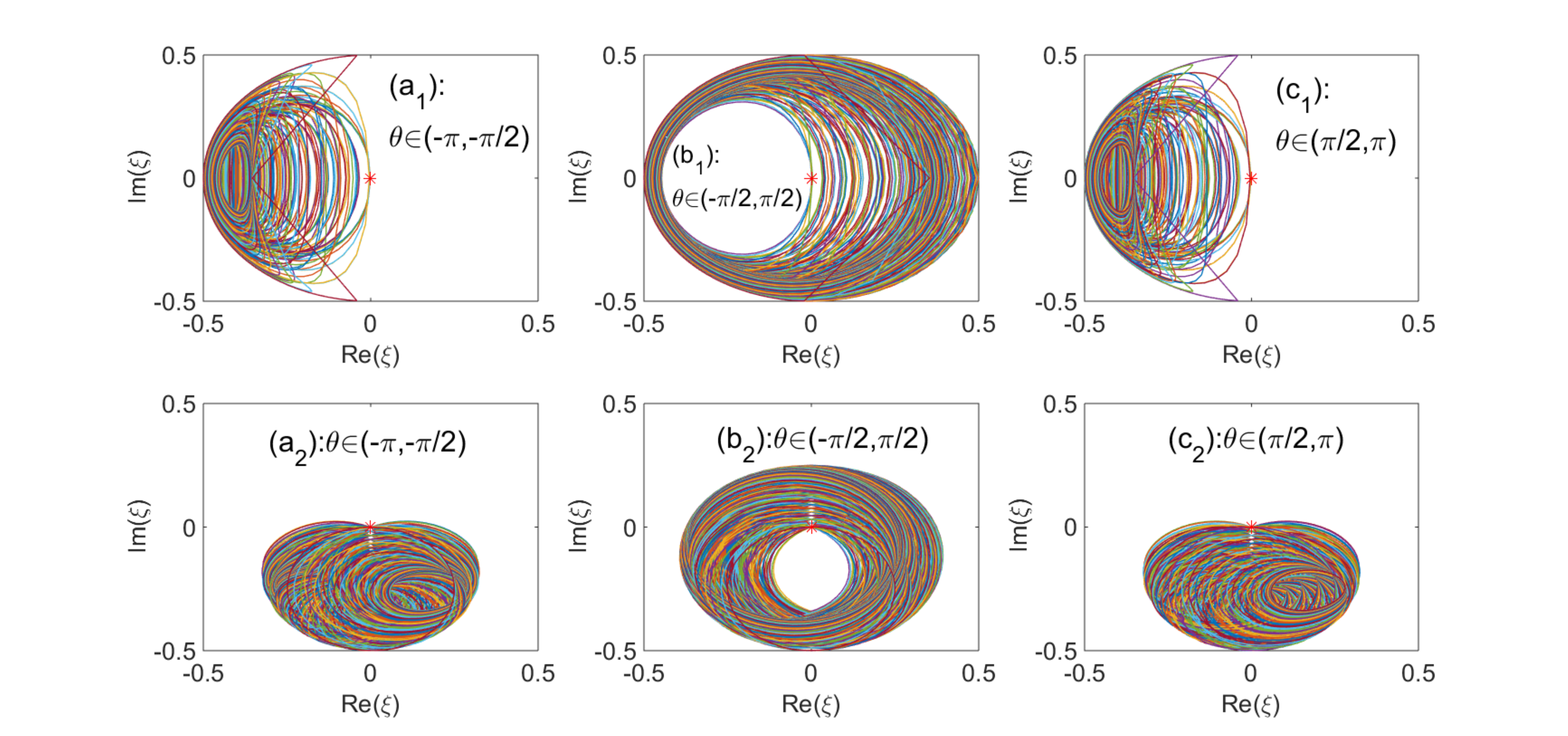}}
\vspace{0cm}
\caption{($a_1$), ($b_1$) and ($c_1$)  are the trajectories of the wind number $\xi$ in the complex plane for the Hermitian case.
	 ($a_2$), ($b_2$) and ($c_2$) are those in the non-Hermitian Hamiltonian case with $\Gamma$=2. The red star is the origin to
	  indicate whether the trajectories of $\xi$ wind around it. $\phi\in[0,2\pi]$ are shown in all these cases. The winding in this figure
	   coincides with the emergence of edge states in Fig.\ref{f:AAHopen} and Fig.\ref{f:AAHopennon}}
\la{f:AAHwind2}
\end{figure*}
\end{center}
In this dimer Hamiltonian, the hopping terms in $H_{odd}$ are intra-cell hoppings and those in $H_{even}$ are the inter-cell hoppings.

The emergence of the edge states with isolated eigenenergies in the energy gap for a system with open boundary condition is a signature 
of nontrivial topological properties corresponding to the nonvanishing topological number in momentum space. Considering the map from the 
one-dimensional version of AAH model to the two-dimensional counterpart, we confirm that the modulation phase $\theta$ is treated
as a momentum component hereafter. Thus the `momentum space' is not puzzled. Next, we consider the topological properties in
terms of $\phi$. And this phase is also treated as a momentum component after Fourier transformation. By transforming the real space
Hamiltonian to the one in momentum space under the periodic boundary assumption, we can examine the topological phases by
topological invariant.

In the basis of $[a^\dag_k\quad b^\dag_k]^T$, the Hamiltonian  in the momentum space can be written as,
\bea
H_{k}&=&\vec{h}\bullet\sigma=h_x \sigma_x+h_y \sigma_y+h_z\sigma_z,\nonumber\\
&&h_x=1-\lambda \cos(\theta)+[1+\lambda \cos(\theta)]\cos(k),\nonumber\\
&&h_y=[1+\lambda \cos(\theta)]\sin(k),\nonumber\\
&&h_z=-v\cos(\theta+\phi)+i \Gamma,
\la{eq:Hamiltoniank}
\eea
where we have added   $-i \Gamma$ to $h_z$ to describe the loss. The dispersion relation by $H^2=E^2\textbf{I}$ is
\bea
E_{\pm}=\pm\sqrt{h_x^2+h_y^2+h_z^2}=\pm\sqrt{A^2+B^2+C-D},\nonumber\\
\label{eq:Eigvalues}
\eea
where $A^2=2(1+\lambda^2\cos^2(\theta))$, $B^2=v^2\cos^2(\theta+\phi)$, $C=2(1-\lambda^2\cos^2(\theta))\cos(k)$
and $D=\Gamma^2+i 2 v \Gamma\cos(\theta+\phi)$.  One of the eigenvectors of $H_k$ with eigenvalue $E_+$ is
\begin{eqnarray}
|u_{k,-h}\rangle =\left(\begin{array}{c}\phi_k^A\\ \phi_k^B \end{array}\right)= \left(\begin{array}{c}\cos\eta \\ \sin\eta \end{array}\right),
 \label{eq:uk}
\end{eqnarray}
where $\eta$ fulfill $\tan(2\eta)=(h_x-i h_y)/h_z$.  Since the gauge does not influence the topological properties, we have
neglected the phase related to gauge.
\begin{center}
	\begin{figure*}[htb!]
		\centering
		\subfigure{\includegraphics[scale=0.4]{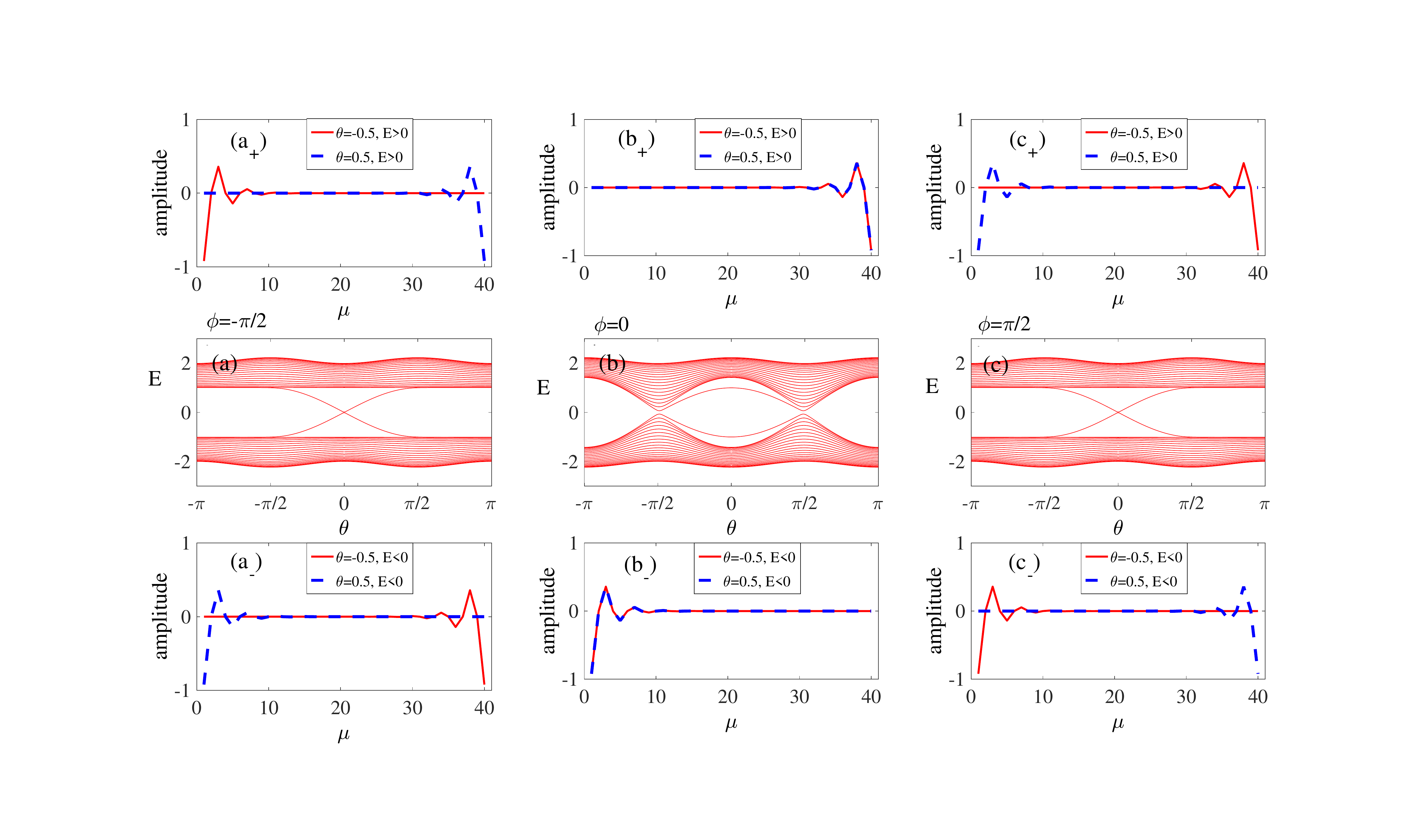}}
		\vspace{-1.5cm}
		\caption{($a_+$), ($b_+$) and ($c_+$) show the amplitudes of the edge states as a function of position for $\phi$=$-\pi/2$, $0$
			 and $\pi/2$ when $\theta$=-0.5 and 0.5 for the positive eigenenergies respectively. ($a_-$), ($b_-$) and ($c_-$) are those
			  for the negative engenenergies. (a), (b) and (c) are the spectrum versus the modulation phase $\theta$ with the parameters
			  $\phi$=$-\pi/2$, $0$ and $\pi/2$ respectively and the other parameters are $\lambda=0.5$, $v=1$ and N=40.}
		\la{f:AAHopen}
	\end{figure*}
\end{center}

Based on the mentioned above, the topological property can be expressed by the wind number which can be defined in different ways 
such as the {\it ratio} of the two components of one of the eigenstates in the momentum representation~\cite{prl102065703}. Additionally, 
the product of the two components:
\bea
\xi=\phi_k^A\phi_k^B=\frac{1}{2}(\bar{h}_x- i\bar{h}_y),
\la{xi}
\eea
can also be used  as a wind number in the complex plane of $(\bar{h}_x,\bar{h}_y)$ where $\bar{h}_x$ and $\bar{h}_y$ are $h_x$
and $h_y$ divided  by $2Re( E)+2i\Gamma$, where $E$ is the eigenenergy and $Re(\bullet)$ returns the real part of $\bullet$. Whether
 the trajectory of $\xi$ wraps the origin  is an signature  of topological property for the system and independent of which eigenstate being used.
 We show the trajectory of $\xi$ in Fig. \ref{f:AAHwind2} for three intervals of $\theta$. To plot  these figures, $\phi\in[0,2\pi]$ have been
randomly examined. It can be seen that although  the trajectories of $\xi$ are different for the Hermitian and non-Hermitian cases, the topological
regions are same in terms of $\theta$. Considering the bulk-boundary correspondence, the nontrivial topological phase corresponds to
the emergence of edge localized states in real space with open boundary condition. To check the topological properties in terms of the
hopping modulation phase $\theta$, we exhibit the energy spectrum of the chain as a function of $\theta$ for $\phi$=$[-\pi/2,0,\pi/2]$
in Fig.~\ref{f:AAHopen}. It can be seen that in both cases the wind number coincide with the emergence of edge states, see
Fig.\ref{f:AAHopen} and Fig.\ref{f:AAHopennon}. Namely, when $\theta\in(-\pi/2,\pi/2)$ for $\phi\in(0,2\pi)$,
the trajectories of $\xi$ wind around the origin, the edge states appear with the energies localized in the energy gap. However in the
presence of non-Hermitian loss, from the real energy spectrum, we can see that the region for the emergence of edge states shrinks.
This may result from that the non-Hermitian loss `draws' the real energy spectrum towards zero which is reflected by the deformation
of the spectrum compared to those in Hermitian case in Fig.\ref{f:AAHopen}. The topological region has not changed
obviously according to the performance of $\xi$. And when $\theta\in(\pi/2,3\pi/2)$, the origin locates outside the trajectory of $\xi$,
thus  edge states do not appear.
\begin{center}
	\begin{figure*}[htb!]
		\centering
		\subfigure{\includegraphics[scale=0.40]{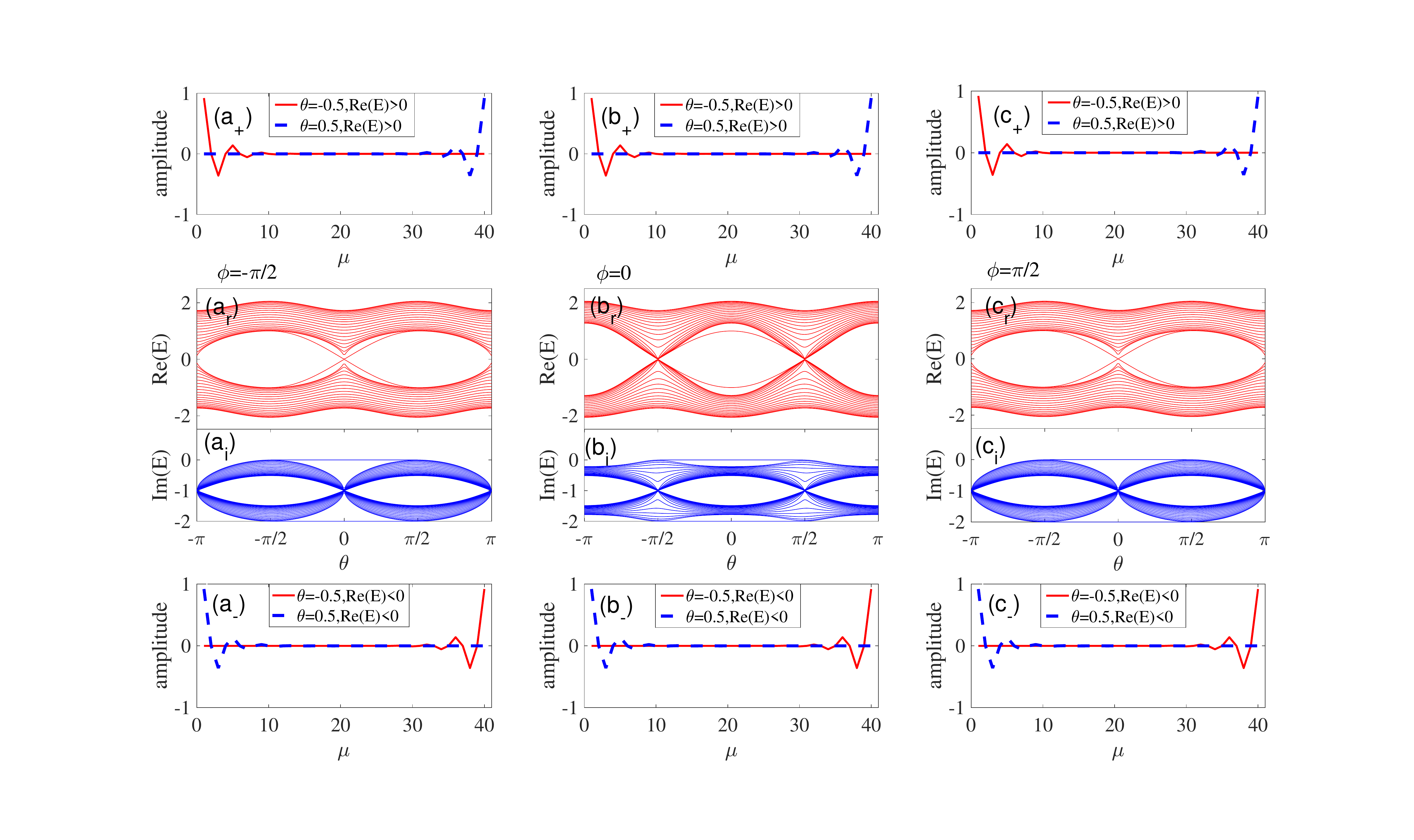}}
		\vspace{-1.5cm}
		\caption{The real and imaginary energy spectra and distributions for the edge states same to those in Fig.\ref{f:AAHopen}
			with loss rate $\Gamma=2$ on B-sites. The other parameters are same to those in Fig.\ref{f:AAHopen}.}
		\la{f:AAHopennon}
	\end{figure*}
\end{center}

Next, we examine about the energy band and topological edge states in more details. It is easy to find  that the edge states appear and their energies
intersect the gap at $\theta_c$=$\pi/2-\phi~(\phi\in[0,\pi])$ and $\theta_c$=$3\pi/2-\phi~(\phi\in[\pi,2\pi])$ which results to $h_z=0$.
The degenerate point results from the particle-hole symmetry of the system,  $\mathcal{S}$: $c_n^{\dag}\rightarrow(-1)^nc_n$
and $c_n\rightarrow(-1)^nc_n^{\dag}$~\cite{prl89077002}. When this symmetry is broken, namely, $h_z\neq0$ here,
the degenerate zero energy edge state vanishes. The population of the edge states are generally localized at the ends of the chain. By
checking the distribution of the wavefunctions for the edge states, we find that for the same $\theta$ or the identical energy $E$, the
two edge states locate at opposite ends of the chain. And for a certain $\theta$, the two edge states locate on A and B-sites respectively.
Since $\partial_{\theta}E$ is the group velocity of the excitation if $\theta$ is regarded as one momentum component in the 2D counterpart,
 one can see that the excitations with opposite directions of velocity locate at the opposite edges. The emergence of these boundary localized
  states manifests that this AAH model belongs to a nontrivial topological phase.
\begin{center}
\begin{figure}
\includegraphics[scale=0.20]{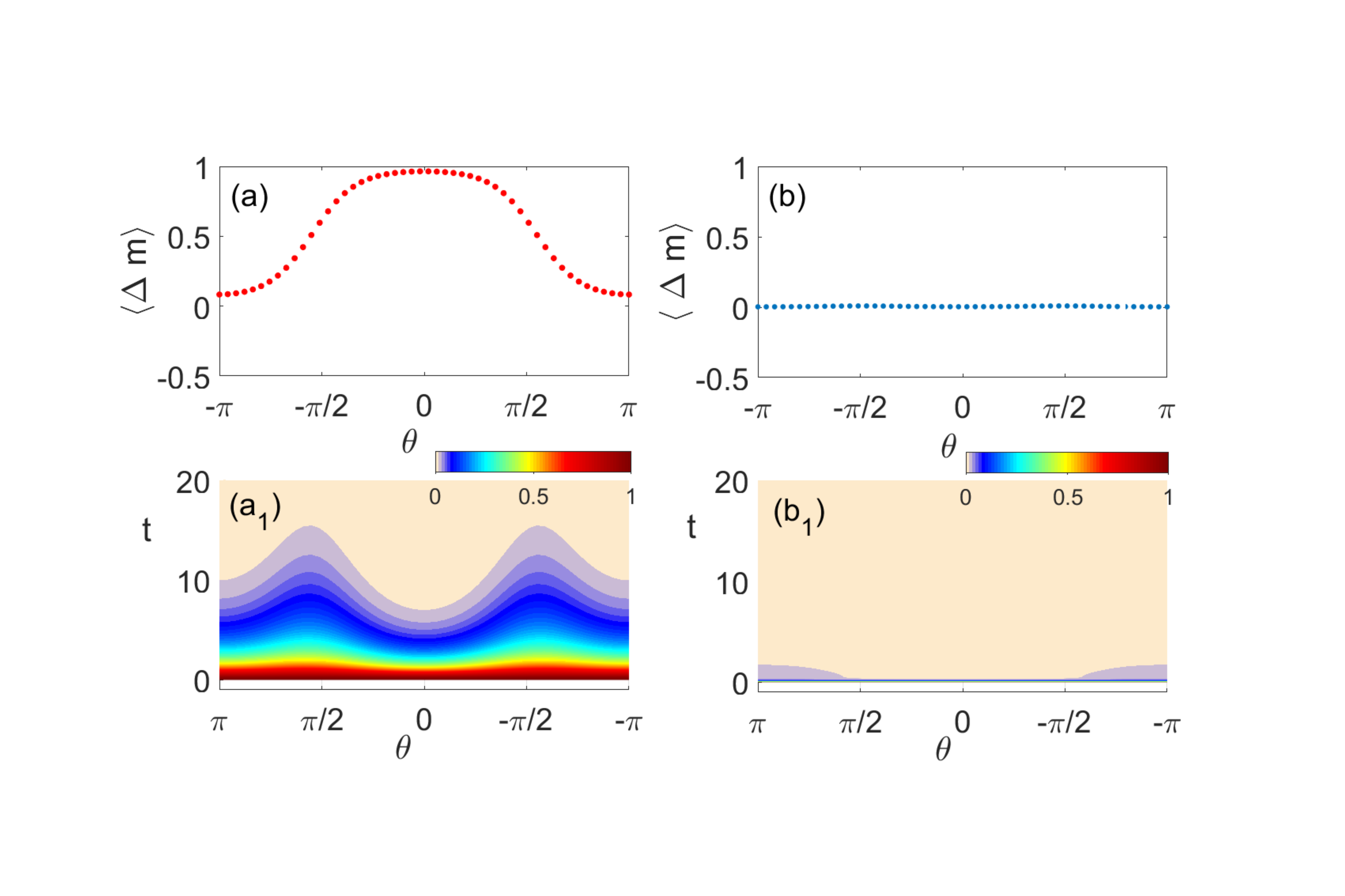}
\vspace{-1cm}
\caption{(a) and (b) are the average displacement versus $\theta$ in the cases of $A$-site  and $B$-site initially excited respectively. ($a_1$)
	 and ($b_1$) are the dynamics of the total excitation $\tau$ versus different $\theta$s corresponding to (a) and (b) respectively. There are 20
	  unit cells (N=40) of $A$-$B$ combination. $\phi=0$ we have set.}
\label{f:AAHdischern}
\end{figure}
\end{center}
\begin{center}
\begin{figure}
\includegraphics[scale=0.22]{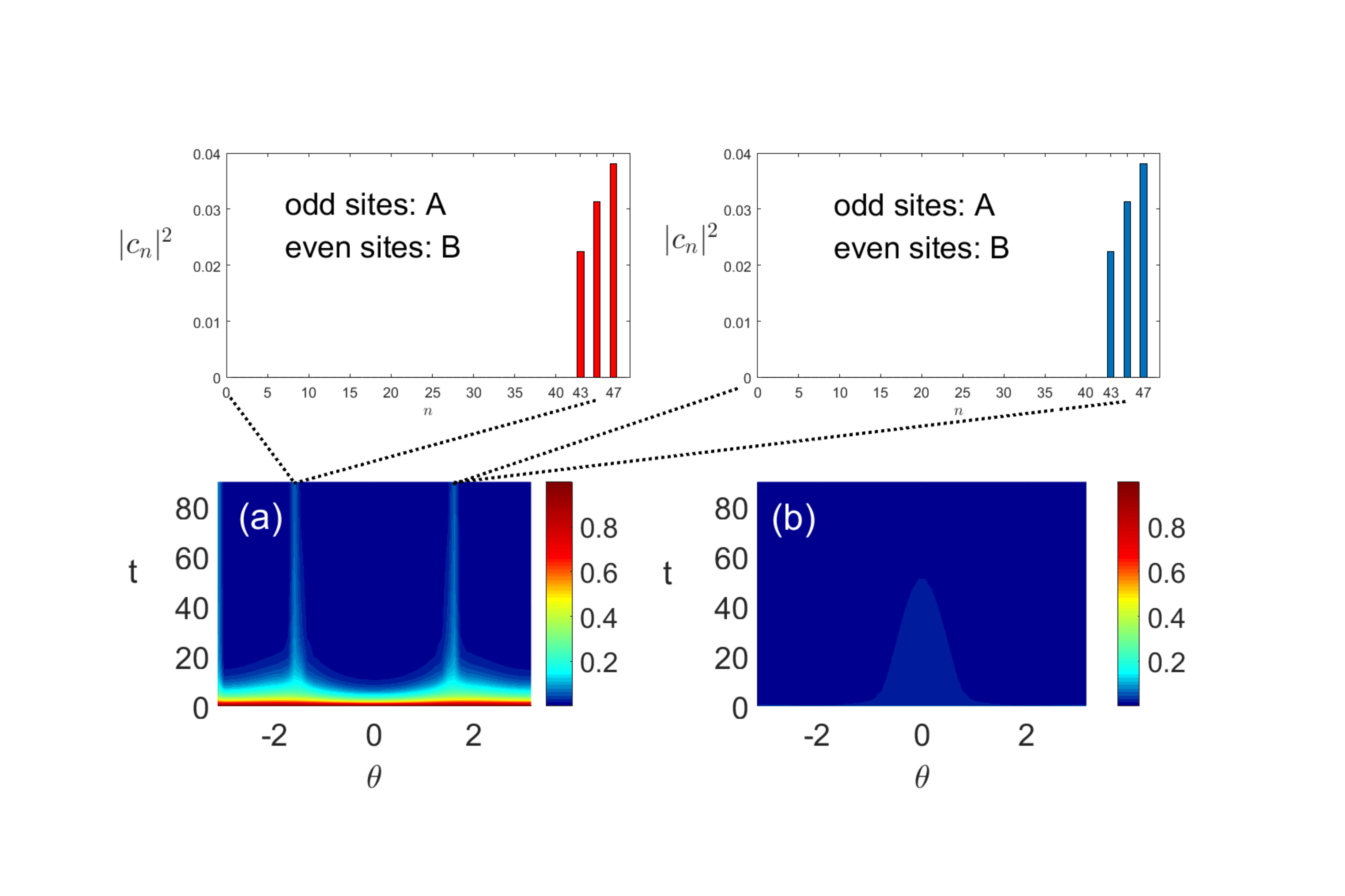}
\vspace{-1cm}
\caption{The evolution of the excitation $\tau$ in (\ref{tau}) for the initial excitation located  on 1-sites (a) and 2-site (b) with the periodic 
	boundary condition. We set  $N=48$, namely there are 24 cells on the ring. The insides of the figure exhibit the distributions of the `dark states'.}
\label{f:AAHdark}
\end{figure}
\end{center}

Compared to the Hermitian case, some intriguing properties may appear in the non-Hermitian case.  Compared to the no-loss case in 
Fig. \ref{f:AAHopen},  we first calculate  the energy spectrum of the chain as a function of  $\theta$  with open boundary condition. 
The results are illustrated  in Fig.\ref{f:AAHopennon}. It can be seen that not only the shape of the energy band changes but also the phase 
of the wavefunction reversed for the same $\theta$.  However, the symmetry  of the wavefunctions remain unchanged  with respect to those 
in the Hermitian case. The edge states still appear in the energy gap in the interval $\theta\in[-\pi/2,\pi/2]$. The imaginary part of the eigenenergy 
is negative indicating decay of the excitation except for slowly decay of the `dark states'. Here the `dark states' are the long-lived states that only 
populate on the non-loss sites which result from the interference of the wavefunction on the chain.  The other witness of topology and the `dark states' 
will be studied in the next section.
\subsection{ Average displacement as a topological index}
In the non-Hermitian case, the mean displacement of the single excitation initially localized on a  non-decay site can serve as a tool to witness the
 topological features of the system~\cite{prl102065703}. It was defined as
\bea
\langle\Delta m\rangle=2\gamma\sum_m(m-m_0)\int_{0}^{\infty}dt|\beta_m(t)|^2,
\la{deltam}
\eea
where $\beta_m$ is the amplitude of the wavefunction on site $B$ in unit cell $m$ at time $t$ and $m_0$ denotes the position for the initial
 single excitation.  For multi-excitation case, further work for $\langle\Delta m\rangle$ may be investigated in the future. Transforming it to momentum space by
 Fourier transformation, the average displacement equals to a wind number of the relative phase between two components of the Bloch
 wavefunction~\cite{prl102065703} and provides a feasible methodology to unraveling the topological properties of the system rather than
  probing the edge states~\cite{science3141757,nature496196,nature452970}.
\begin{center}
	\begin{figure}
		\includegraphics[scale=0.4]{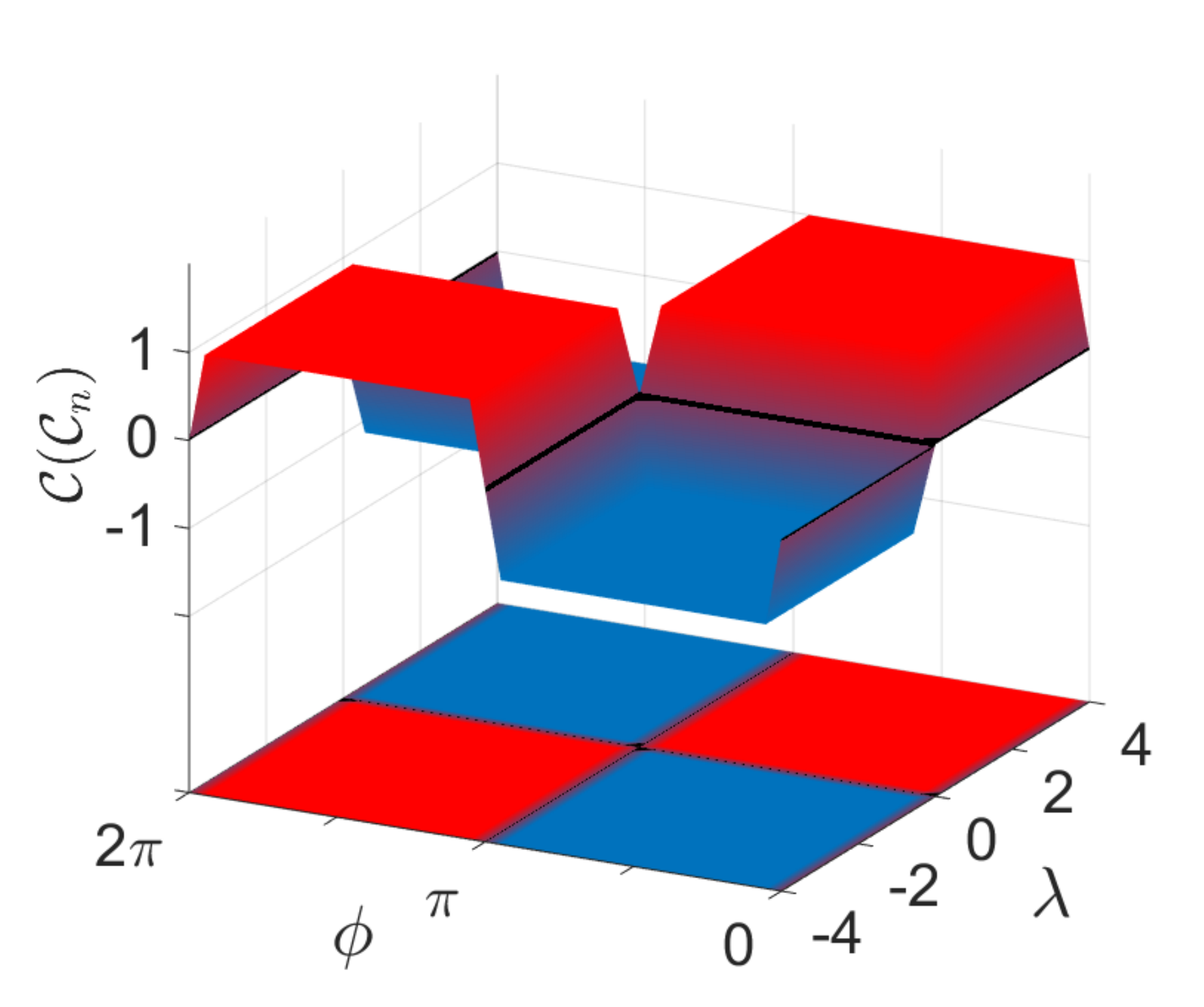}
		\vspace{0cm}
		\caption{The Chern number versus $\phi$ and $\lambda$ in (\ref{chernH}) and it is same for (\ref{chernstate}) versus $\phi$ and $\lambda$ .}
		\label{f:AAHbanChern}
	\end{figure}
\end{center}				
\begin{center}
	\begin{figure*}
		\centering
		\subfigure{\includegraphics[scale=0.45]{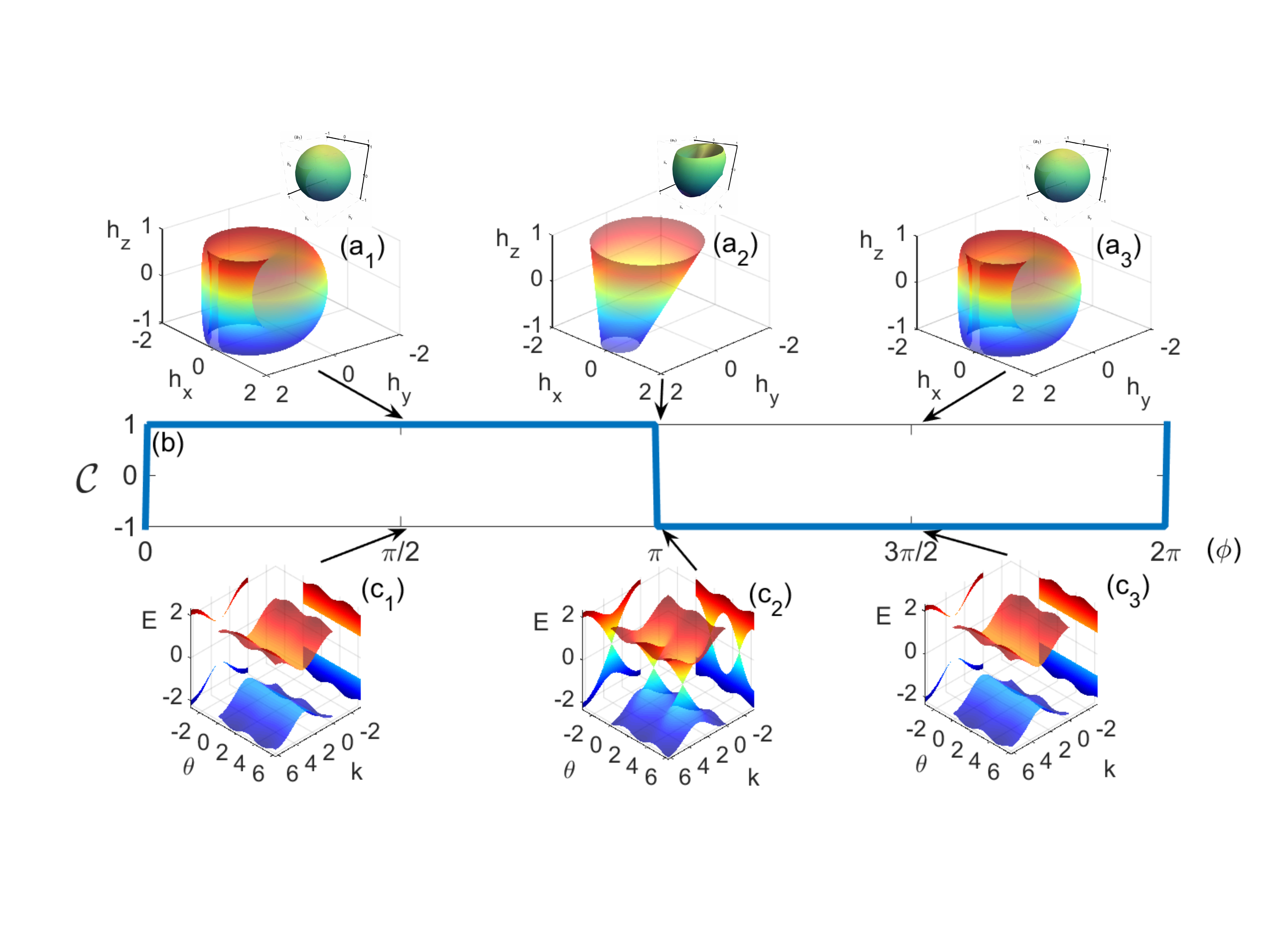}}
		\vspace{-2cm}
		\caption{($a_1$), ($a_2$) and ($a_3$) are the graphical representations of the Chern number in the space of $h_x$-$h_y$-$h_z$
			for $\phi$=$\pi/2$, $\pi$ and $3\pi/2$ respectively. The insets are the normalized graphical representations for the Chern number.
			  ($c_1$), ($c_2$) and ($c_3$)  are the energy spectra corresponding to ($a_1$), ($a_2$) and ($a_3$) respectively. ($b$) exhibits
			  the Chern number defined in (\ref{chernH}). We have set $\lambda$=0.5 and $v$=1.}
		\la{f:AAHchern3D}
	\end{figure*}
\end{center}	

It can be seen that $\langle \Delta m\rangle$ depends on the dynamical history of the wavefunction. The dynamics for the wavefunction is
governed by the equations:
\bea
\dot{\alpha_n}=&&-i(1-\lambda\cos(\theta))\beta_n-i(1+\lambda\cos(\theta))\beta_{n-1}\nonumber\\
&&+i v \cos(\theta_{v})\alpha_n,\nonumber\\
\dot{\beta_n}=&&-i(1-\lambda\cos(\theta))\alpha_n-i(1+\lambda\cos(\theta))\alpha_{n+1}\nonumber\\
&&-i v \cos(\theta_{v})\beta_{n}-2\Gamma\beta_{n},
\eea
where $\alpha_n$ and $\beta_n$ are the amplitudes for the wavefunction on $A$ and $B$ sites respectively. The loss effect resulting from
the minus complex on-site potentials on B-sites can be seen from these dynamical equations. By integrating the dynamical equations, we gain
 the average displacement numerically.

We plot $\langle\Delta m\rangle$ versus $\theta$ in Fig. \ref{f:AAHdischern} (a) and (b).  Compared to the energy spectrum in
Fig. \ref{f:AAHopen}, we can see that  $\theta=\pm\pi/2$ are the transition points for $\langle \Delta m\rangle$ which  coincide with the
appearance of edge state versus $\theta$. And  the total excitation which is defined as
\bea
\tau(t)=\sum_m|\alpha_m(t)|^2+|\beta_m(t)|^2
\label{tau}
\eea
would be longer at the transition points. Then we exhibit the dynamics of the total excitation $\tau$ versus $\theta$ in Fig. \ref{f:AAHdischern}
($a_1$) and ($b_1$). It can be seen that while $\theta=\pm\pi/2$, the lifetime of the excitation is prolonged obviously. This results from the
 existence of `dark state'. To exhibit the `dark states', the distribution of the wavefunction with periodic boundary condition is checked in
 Fig.\ref{f:AAHdark}. It can be seen that there are long-lived states with excitation locating at the non-loss sites, namely, the `dark state'.
 Such states result from the coherence of the wavefunction on the chain when $\theta=\pm\pi/2$. That means with the periodic boundary
  condition, even there is the loss on B-sites, the dark state can survive in a long time scale. However, the coupling between A and B sites
  leads to damping of the excitation gradually.
\section{ topological properties in terms of on-site modulation phase}
\la{ONSITE}
In this section, we study  the effect  of loss on the topological properties in terms of  phase $\phi$ in the on-site modulation. If we
assume $\theta_v=\theta+\phi$,  the system exhibits intriguing topological properties in terms of $\phi$. By transforming the system to momentum
space, the topological properties can be quantified  by Chern number defined by
\bea
\mathcal{C}(\phi)=\frac{1}{4\pi}\int\int_{BZ} d k d \theta(\partial_{k}\vec{h}\times\partial_{\theta}\vec{h}),
\la{chernH}
\eea
where $\vec{h}=(h_x,h_y,h_z)$. From the other  point of view, the  Chern number can be calculated via the  $n$th energy eigenstate, it reads
\bea
\mathcal{C}_n(\phi)=-\frac{1}{2\pi}\int\int_{BZ} d k d \theta(\partial_{k}A_{\theta}^{(n)}\times\partial_{\theta}A_{k}^{(n)}),
\la{chernstate}
\eea
here $A_j^{(n)}(j=k, \theta)$ are the Berry connection given by $A_j^n=-i\langle u^{(n)}|\partial_{j}|u^{n}\rangle$.
These two Chern numbers are identical to describe the topological properties of the system. Later,  we will  take the Chern number
in (\ref{chernH}) as the topological index.

In Fig.\ref{f:AAHbandChern}, we show the Chern number $\mathcal{C}(\mathcal{C}_n)=sign(\lambda\sin(\phi))$ as a function 
of $\phi$ and $\lambda$. With $\lambda=0.5$, we present the topological characters  and the energy spectra versus $\phi$ in Fig .~\ref{f:AAHchern3D}.
It can be seen that $\phi=\pi$ is the critical point for topological phase transition where the energy gap reopens after closing
in the Brillouin zone $\phi\in[0,2\pi]$.  The origin is wrapped by the torus of $h_x-h_y-h_z$ in the Brillouin $\kappa,\theta\in[0,2\pi]$
when $\phi\neq\pi$ where $n=0,~1$. When $\phi=\pi$, the torus of $h_x-h_y-h_z$ degenerate to a closed ribbon across the origin.
The insets of $(a_1)$, $(a_2)$  and $(a_3)$ in Fig.\ref{f:AAHchern3D} show $\bar{h}_i$, $i=x,y,z$  normalized by $|\bar{h}|$. 
When the Chern number is nonvanishing, the sphere of $\bar{h}$ wraps the origin. And correspondingly, the edge states appear in open
 boundary condition in real space. We exhibit the edge spectrum as a function of $\phi$ in the Supplement material by a movie. We can see in the movie 
 that  the edge state appears  except $\phi$=0 or $\pi$ in the Brillouin zone with  $\Gamma$=0. Considering the energy band in momentum 
 space, it can be seen that when $\phi=\pi$, two Dirac points appear. Since the linear dispersion relation near the cones, the excitations 
 with positive and negative energies act like massless particles.

\begin{center}
		\begin{figure}[htb!]
			\includegraphics[scale=0.22]{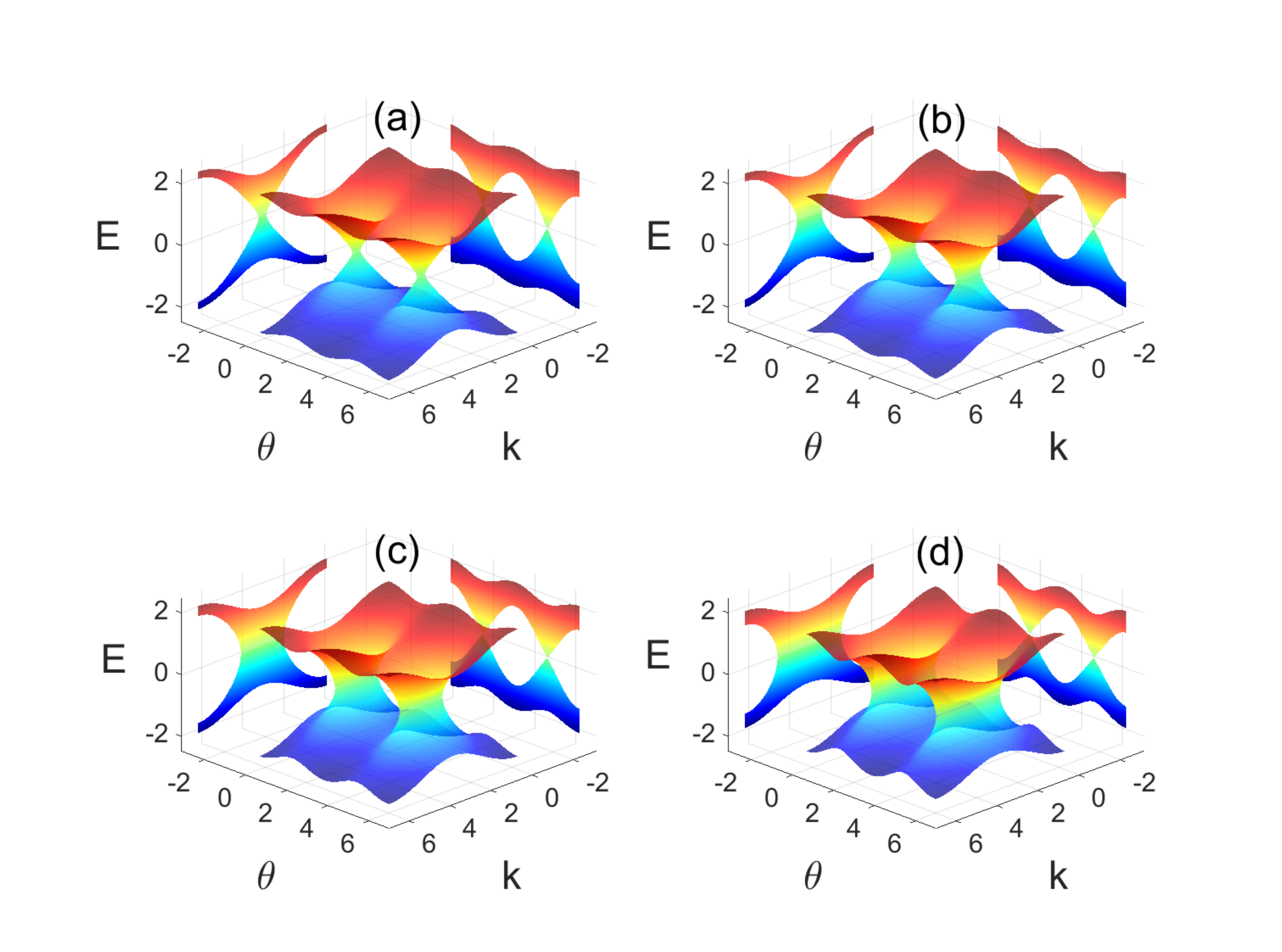}
			\vspace{-0.5cm}
			\caption{(a), (b), (c) and (d) are the real energy spectrum in momentum space for $\Gamma$=0.1, 0.3, 0.5 and 0.7 respectively.
				 $\phi=\pi$ here. The other parameters are the same as those in Fig. \ref{f:AAHchern3D}.}
			\label{f:AAHnon3D}
		\end{figure}
	\end{center}

Next, we examine  the influence of the non-Hermitian loss to the Chern number $\mathcal{C}$, the results are shown  in Fig.\ref{eq:Eigvalues}.
 It can be seen that with the increasing of the loss strength $\Gamma$, the nontrivial topological region is shrunk. Correspondingly, 
 the touching points in the real energy spectrum become lines in the momentum space. In Fig.\ref{f:AAHnon3D}, we exhibit the influence of the
loss to the real energy spectrum. The length of the line of $Re(E)=0$ corresponds to the shrink of the region for the emergence of edge states 
with the open energy gap.

\begin{center}
		\begin{figure}
			\includegraphics[scale=0.25]{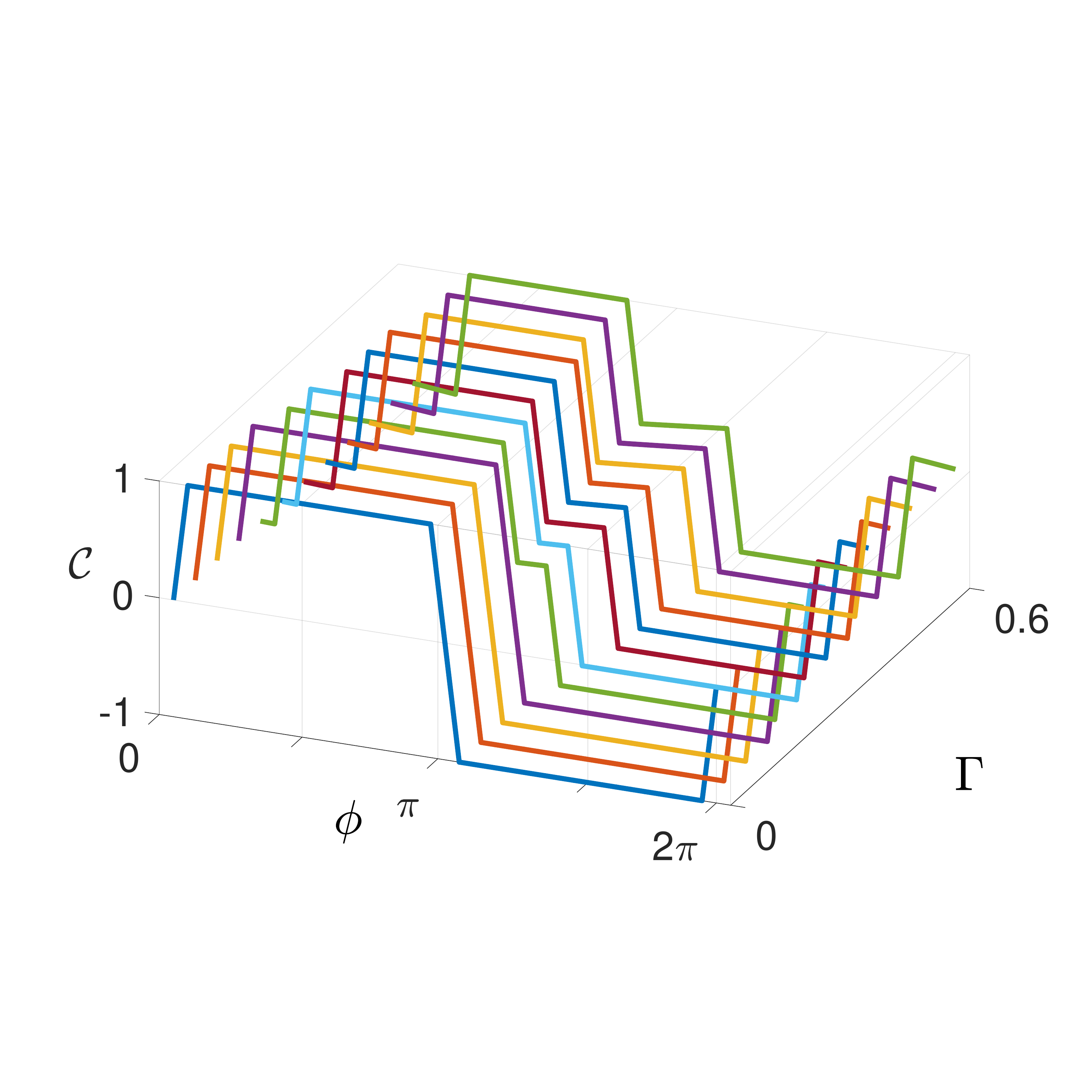}
			\vspace{-1cm}
			\caption{The Chern number versus $\phi$ for a range of non-Hermitian loss strength $\Gamma$ are shown. The
				other parameters are same to those in Fig. \ref{f:AAHchern3D}.}
			\label{f:AAHchernG}
		\end{figure}
	\end{center}

We show the shrink of the region for the emergence of edge states by four different $\Gamma$s in the movie in the Supplement material.
We can see that with increasing of $\Gamma$, the interval of $\phi$ for the band closing points  increase. But the edge states
 do not disappear. Then we conclude that the non-Hermitian loss `draws' the energy spectrum towards zero but dissolve  the
 edge states. The nonvanishing Chern number shown in Fig.~\ref{f:AAHchernG} corresponds to the existence of edge state
 when the energy gap open.
	
\begin{center}
		\begin{figure}[htb!]
			\includegraphics[scale=0.4]{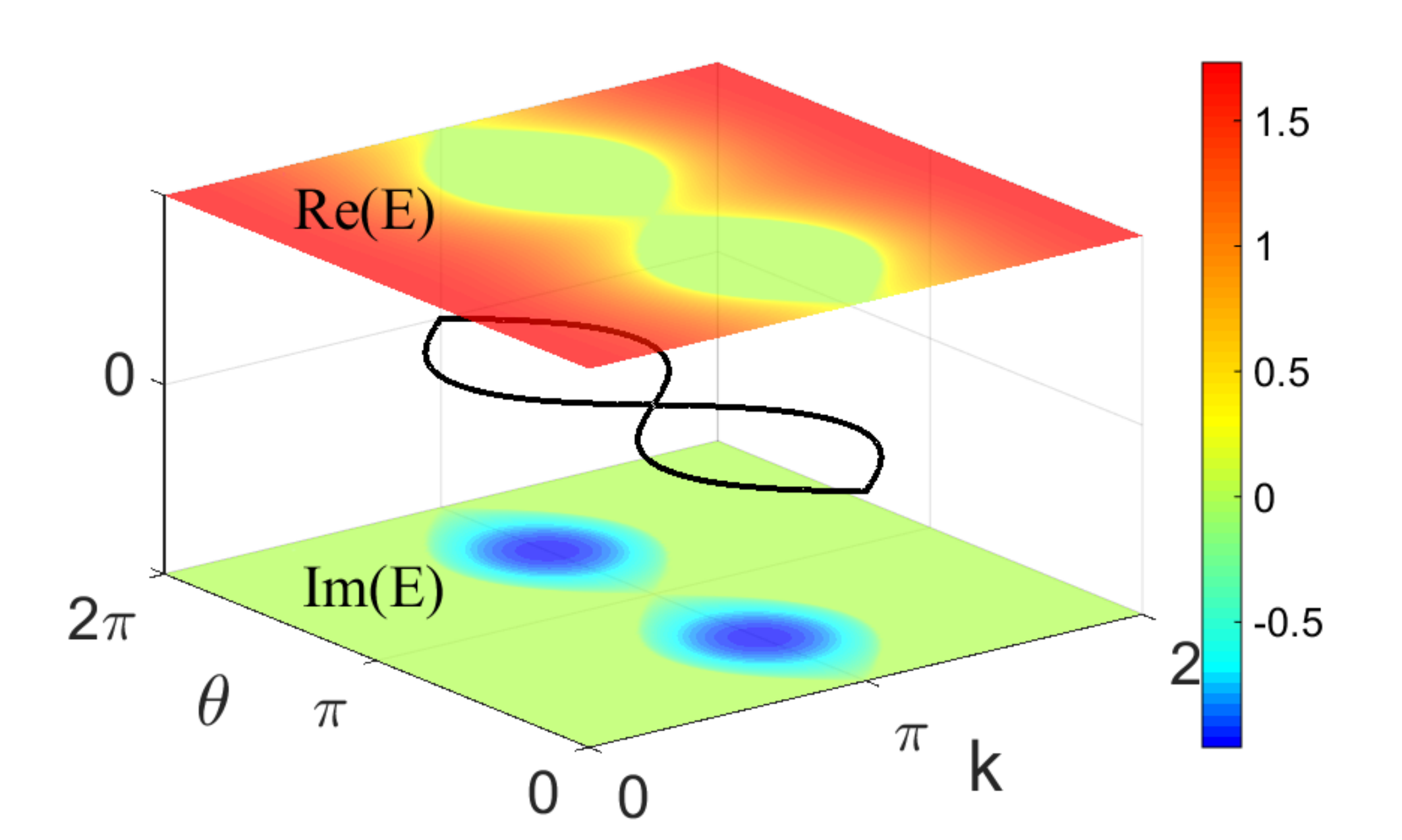}
			\vspace{-0cm}
			\caption{Energy spectrum of the non-Hermitian Hamiltonian and the exceptional points in momentum space. The black loops
				consists of the exceptional points when $\theta_v=\pi/2$.}
			\label{f:AAHexceptpoint}
		\end{figure}
	\end{center}

In the non-Hermitian case, the degenerate points (often called  exceptional points) have interesting properties~\cite{jpa45444016}.
These points are different from the degenerate points in Hermitian case since the eigenstates at these points usually coalesce into one
self-orthogonal state. In the non-Hermitian case, the dispersion relation is $E=\pm(\mathcal{R}+i\mathcal{I})$ where 
$\mathcal{R}=\sqrt{\frac{1}{2}(F+G)}$ and $\mathcal{I}=sign(-v\cos(\theta+\phi))\sqrt{\frac{1}{2}(F-G)}$.  
$F=\sqrt{G^2+4B^4\Gamma^2}$ and  $G=A^2+B^2+C-\Gamma^2$.
We consider the case of $\theta_v=\pi/2$, the energy spectrum $E=\pm\sqrt{G}$. The exceptional points in this case fulfill the
condition: $4\cos^2(k/2)+4\lambda^2\cos^2(\theta)\sin^2(k/2)=\Gamma^2$ in the $k$-$\theta$ space.  We plot the energy spectrum
and the exceptional points in Fig.\ref{f:AAHexceptpoint}. We find that the exceptional points constitute two loops in the parameter
space of $k-\theta$. There are intriguing properties of the exceptional points, e.g., $(k,\theta)=(\pi,\pi)$  is one exceptional point
with the self-orthogonal eigenvector $e^{i \mu}[i\quad 1]^T$ corresponding to the eigenenergy 0 where $\mu$ is an gauge factor.
\section{Robustness of the zero energy edge state}
\label{ROBUST}

\begin{center}
		\begin{figure}[htb!]
			\includegraphics[scale=0.24]{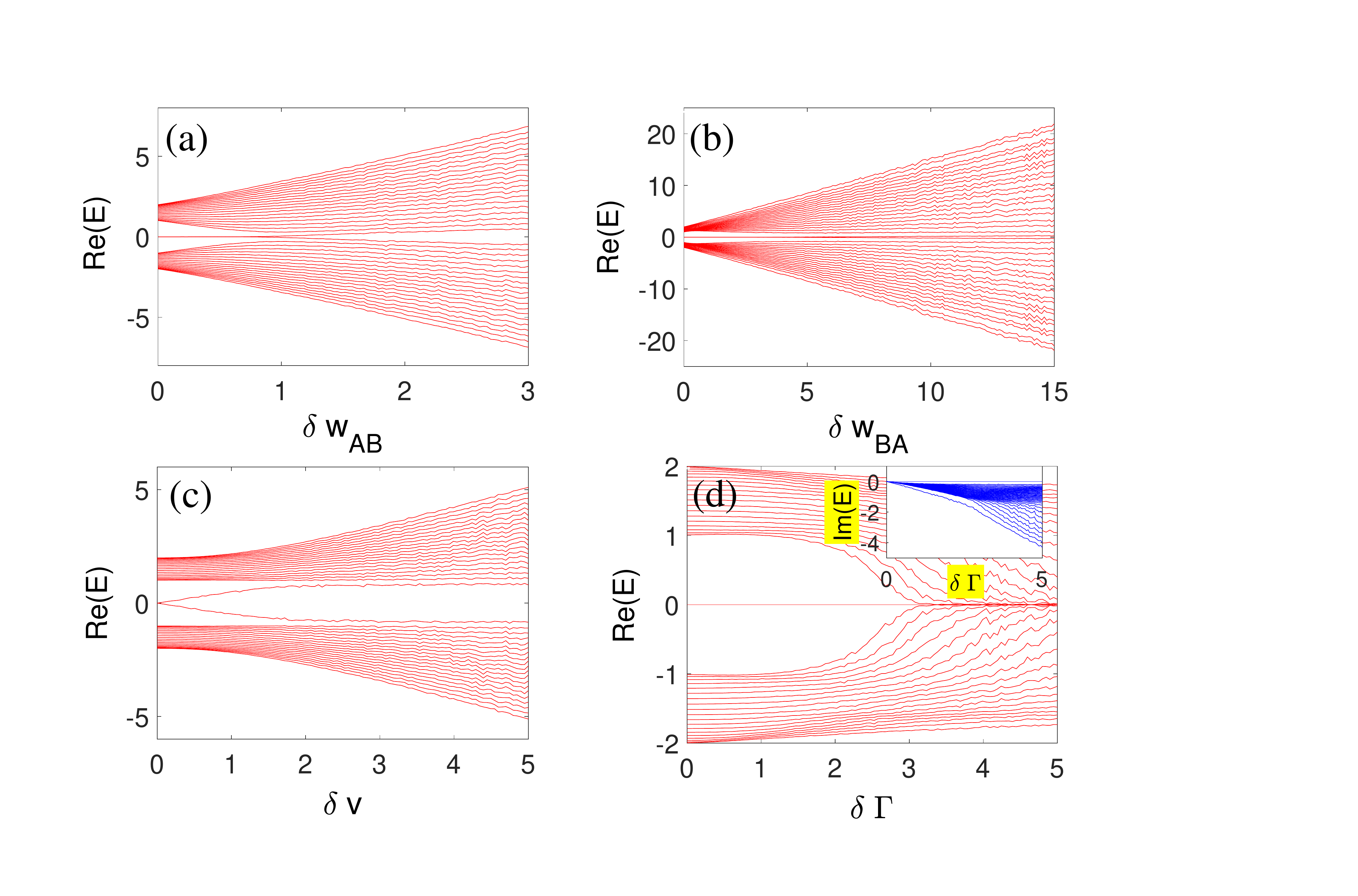}
					\vspace{-1cm}
			\caption{ In (a), (b), (c) and (d): $\delta w_{AB}$ and  $\delta w_{BA}$ are the disorders for the intra-cell and inter-cell hoppings, respectively, 
				 and $\delta v$ and $\delta \Gamma$ are those for the on-site potentials and loss. In each numerical simulation, we assume that the disorders 
				 are randomly distributed on the chain with  strengths uniformly distribute in [0, $\delta x$] ($\delta \Gamma$ locate on B-sites only), here $\delta x$ 
				 are $\delta w_{AB}$, $\delta w_{BA}$, $\delta v$ and $\delta \Gamma$. Here we have set $\theta$=0 and $\phi=\pi/2$ when the zero
				  energy edge state exists. The (non-Hermitian) loss is added to the Hermitian Hamiltonian on B-sites. Each line is an average over 100 simulations.}
			\label{f:AAHdisorder}
		\end{figure}
\end{center}

The existence of the edge states is a nontrivial topological signature in this model. Among the edge states,  the zero energy degenerate states 
are very interesting. In the following by numerical simulations we study the robustness of the zero energy edge state against four kinds of disorders. 
The results are shown  in Fig.\ref{f:AAHdisorder}. From the numerical  simulations, we can see that although the details 
of the robustness of the zero mode against the disorders are different, the states are robust against these disorders but $\delta v$. 
The  zero energy edge state is fragile under the effect of disorders in  $\delta v$ due to the broken  particle-hole symmetry as discussed in 
section~\ref{HOP}. Regardless of the non-zero eigenenergy, the degenerate states are localized in the band gap. With the  increasing of  
 $\delta\Gamma$, the eigenenergies tend to zero. In the previous discussions of this work, we found that with the increasing of the loss, 
 the region for the closing of the bulk band increases. Thus it maybe conclude that the  loss can drive  the real energy band towards zero.  
 So the $E=0$ degenerate eigenvalue is robust against  disorders as long as the particle-hole symmetry is preserved. The disorders 
 of on-site potentials immediately destroy this symmetry which lead  the zero energy edge state split into non-zero states.
	\section{ Experimental setup}\label{SETUP}
A setup to observe  the predicted topological properties can be realized in coupled single mode optical waveguides
 ~\cite{prl109106402,prl103013901,prl115040402,naturephysics7907}. Each waveguide represents a site and the time evolution is 
 equivalent  to the light excitation propagating in the waveguides. The lattice can be fabricated in silica glass by femtosecond direct laser 
 writing technique~\cite{jpb43163001}  or by high resolution  large field e-beam lithography technique on AlGaAs substrate~\cite{prl103013901}. 
 The hopping modulation can be tuned by varying the spacing between the waveguides which provides a way to determine $\lambda$ and $\theta$. 
 The on-site potentials can be modulated by changing the widths of the waveguides which determine $v$ and $\phi$. The loss is introduced by 
 bending the even-waveguides wiggly perpendicular to the plane of the silica glass \cite{prl115040402} in the trigonometrical manner along 
 the propagating direction of the light or by varying the etch depth of the waveguides on AlGaAs substrate. Since the intrinsic loss of the waveguide 
 is identical for all sites, it can be factored out.  And the fluorescence microscopy technique can be employed to observe the light intensity 
 propagating along the waveguides to check the topological properties. Fig~.\ref{f:AAHmodelado} shows a sketch for such a setup.
\section{ Conclusion}
\label{SUM}
In this work, we have studied the influence of  loss to the topological properties of an extended AAH model in terms of hopping and on-site 
modulation phases. We found that the parameter region for the emergence of the edge localized states with open band gap is shrunk in the 
presence of  loss.  We also examine  the average displacement of the single excitation and find that it can witness  the topological properties 
of the system. Long-lived `dark states'  for the chain  are shown in the periodic boundary condition. In terms of on-site modulation phase, 
we found that compared to the Hermitian case, the region for the nontrivial topological phase is also shrunk in the presence of  loss. And the 
zero-energy edge states are robust against intra-cell, inter-cell and non-Hermitian loss disorders but fragile against those  in the on-site 
potentials since the particle-hole symmetry is broken in the last case. The energy spectrum tends towards zero when the  loss disorders increase. 
Finally, we propose an experimental setup based on coupled waveguides to implement this model.

\section*{ACKNOWLEDGMENTS}
We thank Prof. L. C. Wang at Dalian University of Technology for valuable discussions.
This work is supported by the National Natural Science Foundation of China (Grant No. 11534002 and 61475033).

\end{document}